# Electrostatic deflection of the water molecule, a fundamental asymmetric rotor

Ramiro Moro,[1] Jaap Bulthuis,[2] Jonathon Heinrich,[3] and Vitaly V. Kresin[3]

[1]*Department of Physical Sciences, Cameron University,*
*Lawton, Oklahoma 73505, USA*

[2]*Department of Physical Chemistry and Laser Centre, Vrije Universiteit,*
*De Boelelaan 1083, 1081 HV Amsterdam, The Netherlands*

[3]*Department of Physics and Astronomy, University of Southern California,*
*Los Angeles, California 90089-0484, USA*

An inhomogeneous electric field is used to study the deflection of a supersonic beam of water molecules. The deflection profiles show strong broadening accompanied by a small net displacement towards higher electric fields. The profiles are in excellent agreement with a calculation of rotational Stark shifts. The molecular rotational temperature being the only adjustable parameter, beam deflection is found to offer an accurate and practical means of determining this quantity. A pair of especially strongly responding rotational sublevels, adding up to $\approx 25\%$ of the total beam intensity, are readily separated by deflection, making them potentially useful for further electrostatic manipulation.

# I. INTRODUCTION

It may be appropriate to regard H$_2$O as the most basic asymmetric top molecule, in view of its general importance, clear structure, high dipole moment ($p$=1.855 D [1]), low mass, and very large rotational constants ($A$=27.33 cm$^{-1}$, $B$=14.58 cm$^{-1}$, and $C$=9.50 cm$^{-1}$ [2]: the highest set of values for any stable polyatomic molecule).

Surprisingly, it appears that the only experiment to measure its deflection in a molecular beam took place in 1939 [3]. The result was only a qualitative observation of field-induced beam broadening and a rough estimate of the dipole magnitude. At that time, the Stark deflection patterns of linear rotators and symmetric tops already had been worked out, but the asymmetric rotor was considered "almost impossible to handle. […] Hence no quantitative approach to the study of the electrical nature of asymmetrical top molecules by the beam method is practicable" [4].

The situation having evolved, this paper describes a measurement of the deflection of a beam of water molecules by an inhomogeneous electric field and its concordance with a calculation of the rotational Stark shifts. The experimental arrangement is outlined in Sec. II and the calculation of electrostatic deflections in Sec. III. Sec. IV presents the results of the measurement, compares them with the theory, and discusses the implications.

# II. EXPERIMENTAL ARRANGEMENT

The beam deflection apparatus has been described elsewhere [5]. A supersonic molecular jet is formed by expansion of water vapor, either neat or mixed with He carrier gas, from a stainless steel oven into vacuum through a 75 μm diameter nozzle. The beam passes through a skimmer into the second vacuum chamber, where it is collimated by a 0.25 mm × 2.5 mm slit and passes through a "two-wire" inhomogeneous electric field [6,created by a pair of metal plates of length $l$=152 mm and separated by a 2.49 mm gap. By applying voltage of up to 25 kV between the plates, electric fields to 80 kV/cm and field gradients to 380 V/cm$^2$ were created.

The deflected molecules then pass through a rotating beam chopper and travel to a quadrupole mass spectrometer (UTI-100) where they are ionized by electron bombardment at 70 eV, mass selected ($H_2O$ registers in the mass spectrum mainly as $H_2O^+$ [8]) and detected by an analog multiplier. The detector output is fed into a lock-in amplifier together with the chopper synchronization pulses. The purpose is two-fold: to eliminate noise from the background water in the chamber and to determine the beam velocity $v$ from the phase difference between the chopping and detection signals. The molecular velocities were in the range of 800 m/s to 1300 m/s depending on the chosen source conditions. The velocity spread of a supersonic beam is low, therefore $v$ can be treated as a constant for all molecules in a particular experiment.

To determine the beam profiles, a 0.25 mm wide slit is positioned in front of the detector entrance, a distance $L$=712 mm past the middle of the deflection field plates. The slit is scanned across the beam by a stepper motor in steps of 0.25 mm, measuring the intensity at each position with and without voltage applied to the deflection plates.

The direction of the deflecting electric field $E$ and its gradient will be labeled the $z$-axis. The net deviation $d$ of an individual molecule is proportional (1) to the ratio of the impulse $F\Delta t$ received from the field to the original forward momentum $m_{H_2O} v$, and (2) to the flight time from the field region to the detector slit. The time spent between the deflection plates is $\Delta t = l/v$, and the flight time to the detector is $L/v$. For a molecule in an eigenstate $i$ adiabatically traversing the electric field region, the force in the field direction ($z$) is given by the shift of the energy level ($\varepsilon_i$): $F_i = \partial \varepsilon_i / \partial z = (\partial \varepsilon_i / \partial E) \cdot (\partial E / \partial z)$. For a fixed plate geometry, $\partial E / \partial z \propto E$, and therefore for the deviation of this molecule we can write, subsuming all geometrical constants into one instrumental factor $C$,

$$d_i = C \frac{E}{m_{H_2O} v^2} \cdot \left( \frac{\partial \varepsilon_i}{\partial E} \right). \quad (1)$$

The factor $C$ was determined from calibration measurements on a beam of Ar atoms and checked with $SF_6$, agreeing with the latter's standard polarizability within 2.3% [5].

The slopes $\partial \varepsilon_i / \partial E$ can equivalently be called the effective dipole projections on the $z$-axis, since $\Delta \varepsilon = \vec{p} \cdot \vec{E}$. They were derived from a calculation of the Stark diagram of $H_2O$'s rotational states, as described in the next section, plus a small electronic dipole polarizability contribution.

### III. CALCULATION OF MOLECULAR DEVIATION BY STARK FORCES

#### A. Evaluation of Stark curves

The Stark shifts of the rotational states of $H_2O$ and $D_2O$ were calculated on the basis of the Hamiltonian

$$H = A J_a^2 + B J_b^2 + C J_c^2 + pE \cos \theta, \quad (2)$$

where $J_a$, $J_b$ and $J_c$ are the components of the angular momentum along the three principal axes of inertia, with corresponding rotational constants $A$, $B$ and $C$. The last term on the right-hand side is the (Stark) interaction of the molecular dipole with the electric field $E$. Since it describes a rigid asymmetric rotor, Eq. (2) would in fact be inadequate for an accurate description of the water molecule's rotational spectrum. In particular, centrifugal distortions, which are large for water, should be taken into account. However, as will be shown below, Eq. (2) is sufficient for the present purpose of calculating the deflection patterns.

For $H_2O$ we used the dipole moment and equilibrium rotational constant values listed in the Introduction; for $D_2O$ the following values were used: $A$=15.39 cm$^{-1}$, $B$=7.26 cm$^{-1}$, and $C$=4.85 cm$^{-1}$ [9] and $p$=1.87 D [10].

Following Ref. [11], the eigenvalues of Eq. (2) are found by using the set of symmetrized symmetric-rotor eigenfunctions $|JKM>$ and truncating the Hamiltonian matrix at sufficiently high values of $J$ and $K$. Normally, $J$ and $K$ values up to 24 were included, which, given the large rotational constants of $H_2O$ as well as $D_2O$, is more than sufficient for the range of energies of interest. $M$ is the only good quantum number, and the diagonalization of the truncated matrix is done for all $M$-values that contribute detectably to the beam population. (Absolute values of $M$ were taken, because positive and negative $M$ values occur in pairwise degenerate eigenfunctions. This two-fold degeneracy was taken into account when assigning the statistical weights.) The maximum value of $|M|$ was set to 10 for $H_2O$ and to 12 for $D_2O$.

In the case of water, it is convenient to take the molecular $b$-axis as the $z$-axis. Then $K$ is the projection of $J$ on the dipole axis. With this choice, the Stark Hamiltonian takes its simplest form and does not mix states with different $K$. The rotational Hamiltonian mixes $K$ with $K\pm2$ states, so that matrix elements with odd and even $K$ can be treated as separate blocks. This reduces the size of the Hamiltonian matrix, but more importantly, it simplifies the assignment of spin statistical weights [12] and, moreover, avoided crossings of the Stark curves can easily be tracked down. In the conventional notation of asymmetric-top eigenfunctions, $K_-$ and $K_+$ (denoting the prolate and the oblate top limits, respectively) are used as subscripts for *J*, or alternatively the states are labeled by *J*, $\tau = K_- - K_+$, and *M*. We follow the latter notation.

Comparing the rotational energies [the eigenvalues of Eq. (2) for $H_2O$ at zero electric field] with the experimental values given in Ref. [13], we find that for levels up to about 1000 cm$^{-1}$ the deviations are ≤2%, while for higher-energy levels (especially those with large $K_a$) stronger deviations occur. The majority of the calculated energies lie above the corresponding experimental ones. If the Stark shifts may be described by second-order perturbations, which is a reasonable approximation for water, the shifts and the slopes of the Stark curves have about the same error as the energies, or even less if the errors in the interacting energy levels have the same sign. Since the low-*J* states dominate the deflection patterns by virtue of their Boltzmann weights, and since the rotational temperature derived from the deflection pattern has an uncertainty of about 8% (see below), the treatment of water as a rigid rotor is fully justified in the present circumstances.

The calculated Stark curves for the lowest states of $H_2O$ are displayed in Fig. 1. No complications due to avoided crossings take place. The Stark curve slopes $\partial \varepsilon_i / \partial E$, which determine the force on the molecules, are calculated numerically by incrementing the field by 0.01 kV/cm. From the curves it is immediately visible that molecules in states $|J\tau M\rangle = |1,-1,1\rangle$ and $|1,1,1\rangle$ will undergo large and opposite deflections. Both have a spin-statistical weight of 3, compared to 1 for the $\tau$ even states. These states are indeed prominent in the deflection profiles, see Sec. IVB.

For $D_2O$ the Stark curves are very similar, although the energy differences between levels are, of course, smaller. However, the spin statistical weights are markedly different: the weight of $\tau$ even states is double that of the $\tau$ odd states. For this reason, the deflections with the highest intensities generally do not originate from the same states as for $H_2O$.

**B. Distribution of deflection probabilities**

Once the energies and slopes of the rotational Stark levels are calculated, they can be employed to evaluate the deflection pattern of molecules in the beam. For a given value of the applied field, the Stark sublevels are populated with Boltzmann factors according to an assumed rotational temperature $T_{rot}$, and their slopes at that point are used to calculate the corresponding deflections according to Eq. (1). A small (~2%) contribution is added in the form of the static electronic polarizability of $H_2O$ ($\alpha$=1.47 Å$^3$ [14]), which leads to an additional uniform shift $\partial\varepsilon_i/\partial E$=-$\alpha E$.

The result can be depicted as a stick probability distribution of deviations, as illustrated in Fig. 2. The height of each stick represents the probability of a water molecule occupying a certain rotational Stark sublevel, and thus being deflected to the corresponding position in the detecting plane. Equivalently, it is the deviation pattern that would be produced if an infinitely thin beam of water molecules at a certain $T_{rot}$ passed through our deflection plates. As will be described in Sec. IV, for comparison with the experimental data this pattern is convolved with the real beam profile and the rotational temperature remains the only adjustable parameter.

Fig. 3 plots the average projection of the molecular dipole on the field axis, which represents the small average overall deflection of the molecular ensemble (i.e., the first moment, or the "center of mass" of the stick distribution). It shows that for rotational temperatures above ≈100 K and over a wide range of field magnitudes, this quantity is close to

$$\langle p_z \rangle = \frac{p^2 E}{3 k_B T_{rot}} \qquad (3)$$

In other words, the "orientation cosine" of the average angle between the molecular dipole axis and the field direction is well fitted by ⟨cosθ⟩=⟨$p_z$⟩/$p$=$pE$/(3$k_B T_{rot}$). This is the same form as one would obtain from the statistical Langevin-Debye orientational susceptibility formula [15], however, the present situation is different: the molecules are not in thermal equilibrium in the field but instead respond to adiabatic entry into the electric field region. Although the fact that low-field response should scale with the ratio $pE/T_{rot}$ is expected on dimensional grounds [16], the value of the numerical coefficient is not universal: sample calculations for linear and symmetric rotors demonstrate that it can depend on this ratio as well as on the ratio of the rotational constants. Some precedents are discussed, e.g., in Ref. [17]. It is intriguing that the asymmetric water molecule, treated as a rigid rotor, follows the Langevin-Debye average so closely, even at relatively low temperatures (the rotational constants of $H_2O$ are equivalent to 40 K, 21 K, and 13 K) and without external perturbations to the rotational motion [18,19].

## IV. RESULTS AND DISCUSSION

### A. Comparison of experiment and theory. Determination of the rotational temperature

To compare the calculated deflection probability patterns such as those in Fig.2 with the experimental picture, we need to convolute the former with the actual beam shape, i.e., with the undeflected intensity profile measured by the scanning slit. This profile is shown as the thin solid black line in Fig. 4. This experimentally obtained function is then convoluted with the stick probability distributions calculated for various values of the electric fields, and the results are shown in Fig. 4 as the red dashed lines. Finally, the experimentally measured deflection profiles are shown as thick red solid lines. Within the central region subtended by the mass spectrometer entrance area the agreement is extremely satisfying.

As mentioned in Sec. III B, the only adjustable parameter in the above procedure is the rotational temperature. Fig. 5 illustrates the tight constraint on $T_{rot}$ produced by fitting the profiles in Fig.4. Put another way, the molecular deflection technique offers a practical and accurate method of determining the supersonic beam's rotational temperature.

The calculations and measurements were also carried out with a beam of heavy water molecules, $D_2O$, under different expansion conditions. Fig. 6 shows that the results were slightly less precise, but satisfactory.

### B. Electrostatic separation of individual high-susceptibility quantum states

The stick distributions in Figs. 2 and 4 depict two prominent peaks which evolve to very large deviations in both directions. As mentioned above, the Stark level diagram in Fig. 1 reveals that they come from two specific quantum states, $|J\tau M\rangle=|1,\pm1,1\rangle$ which display exceptionally steep Stark shifts and are strongly populated at the rotational temperature of this experiment. The peaks' strong deflections carried them beyond the acceptance window of our detector when the latter was focused on the central part of the beam, as in Fig.4. In order to verify that these predicted satellites were indeed present and intense, we mounted the mass spectrometer 2.7 mm off-axis, towards the expected position of one of them. Fig. 7, a combined plot, shows that an intense bump is indeed clearly observed at the prescribed location.

These particular molecules are low-field seekers (hence their negative deflection), and obviously a companion high-field seeker signal will be present at the positive deflection position.

As emphasized above, both of these "beamlets" are made up of molecules in the individual Stark sublevel of a single (low-energy, hence high-occupancy) rotational quantum state. The superposition of their intensity, very strong response to the electric field, and the lightness of the $H_2O$ molecule may make these state-selected species appealing subjects for manipulation by electrostatic forces, such as molecular Stark deceleration [20] and electrostatic steering [21].

## ACKNOWLEDGMENTS

We would like to thank Drs. Philippe Dugourd, Bretislav Friedrich, and Maxim Olshanii for useful discussions, and Roman Rabinovitch and Chunlei Xia for their help in setting up the experiment. J.B. acknowledges stimulating discussions on rotational Stark effects of water with Drs. Pepijn Pinkse and Sadiq Ringwala. V.K. is grateful to the J. Heyrovský Institute of Physical Chemistry, Academy of Sciences of the Czech Republic, for its hospitality during the preparation of this paper. This work was supported by the U.S. National Science Foundation under grant No. PHY-0354834.

# FIGURE CAPTIONS

**Fig. 1.** Stark curves for the lowest states of $H_2O$. The states are represented by $J$, $\tau$, and $M$. The spin statistical weights for states with $\tau$ odd are three times as large than those for $\tau$ even.

**Fig. 2.** Deviation probability distributions for the beam of $H_2O$ molecules, calculated from their Stark shifts and electronic polarizability. The parameters for this example are: $v$=1250 m/s, $E$=63.5 kV/cm, $\partial E/\partial z$=300 kV/cm$^2$, and $T_{rot}$=30 K (a) or $T_{rot}$=300 K (b).

**Fig. 3.** A comparison between the calculated average projection of the dipole moment of the water molecule in a thermal bath (Langevin function [15], solid lines) and isolated in a beam (Stark shift calculation, dashed lines) for two magnitudes of the external electric field. For $T_{rot}$>100 K, the difference is below 10%.

**Fig. 4.** Comparison of the experimental deflections of $H_2O$ molecules with the patterns derived from rotational Stark calculations. The supersonic beam was produced with the source reservoir temperature of 50°C, nozzle temperature of 65°C (nozzle diameter 75 μm), and helium carrier gas pressure of 670 Torr. The thin solid black lines are the experimental undeflected beam profiles. The solid thick red lines are the experimental profiles taken with six different voltages applied to the deflection plates. The blue stick patterns are the deviation probability distributions (as in Fig. 2) calculated for the beam at a rotational temperature of 84 K, the best fit value from Fig. 5. The dashed red lines show the expected deflection profiles obtained by convolving the theoretical deviation probability distributions with the undeflected peak profiles. Notice that in this experimental configuration the beam can be detected only within the mass spectrometer acceptance window, indicated by the vertical margins.

**Fig. 5.** Goodness of fit of the deflection profiles as a function of $H_2O$ rotational temperature for the six experiments shown in Fig. 4; the best estimate is $T_{rot}$=84±8 K.

**Fig. 6.** Experimental and theoretical profiles for heavy water ($D_2O$) molecules prepared with the source and nozzle at room temperature, without carrier gas. Lines are marked in the same way as in Fig. 4, and the best-fit rotational temperature is 167 K.

**Fig. 7.** Detection of the intense low-field seeker sideband for $H_2O$, at a setting of $V$=20 kV, $E$=63.5 kV/cm, $\partial E/\partial z$=300 kV/cm$^2$. The central profile, from Fig. 4, was extended by shifting the mass spectrometer off-axis, thereby detecting the predicted high-deflection "beamlet" shown as the solid thick black line. Its position and strength (11% of the total intensity) are reproduced without any adjustable parameters (dashed red curve). A companion high-field seeker sideband of similar magnitude (14%) will be present on the opposite side.

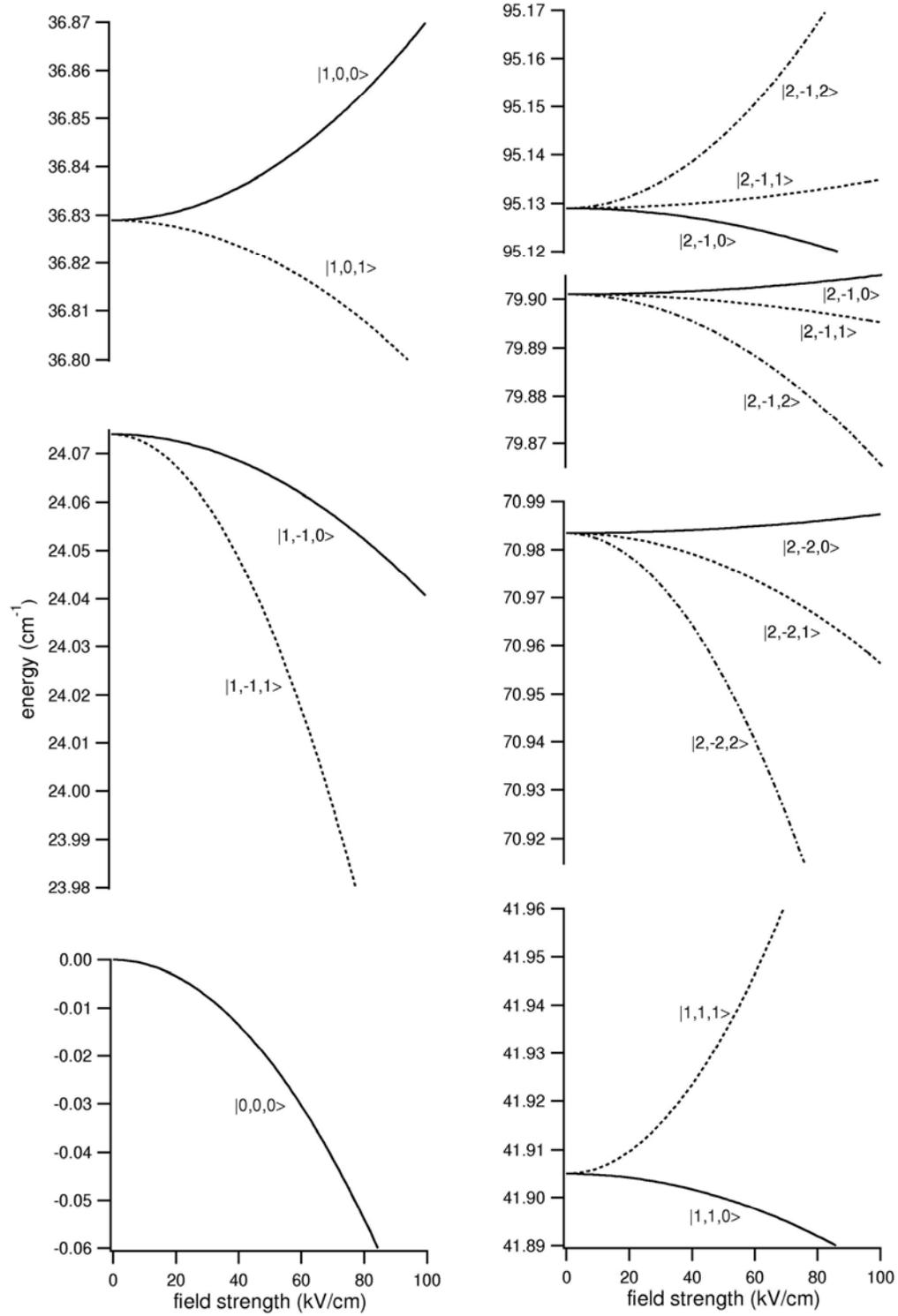

**Figure 1**

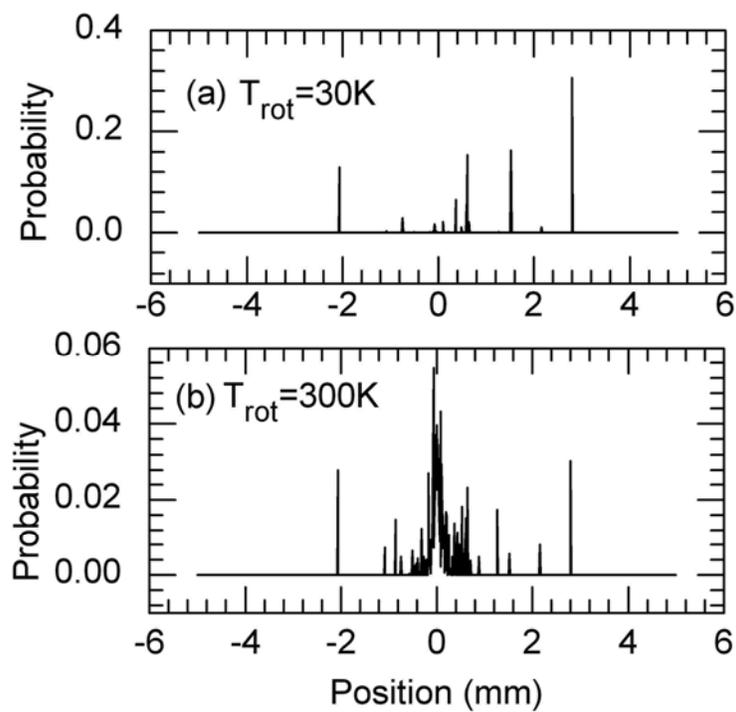

**Figure 2**

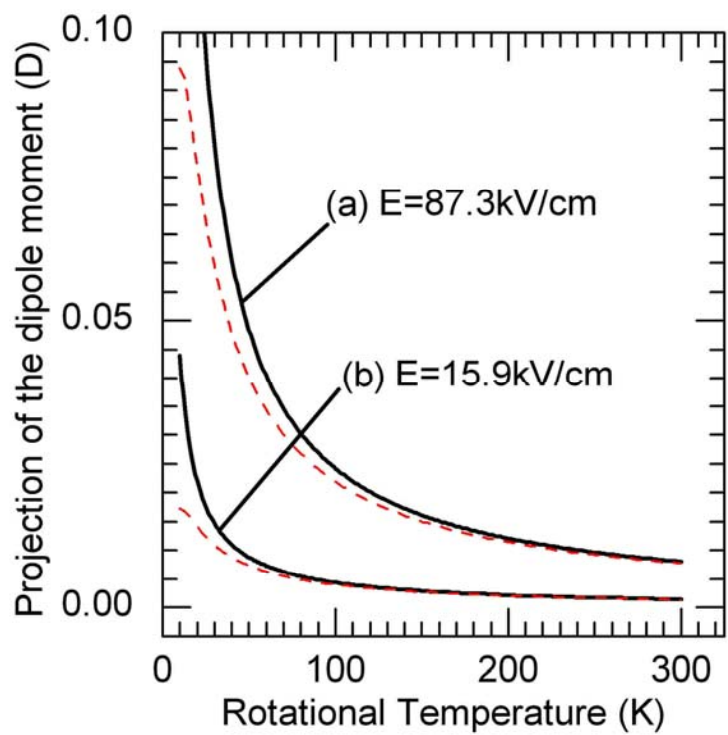

**Figure 3**

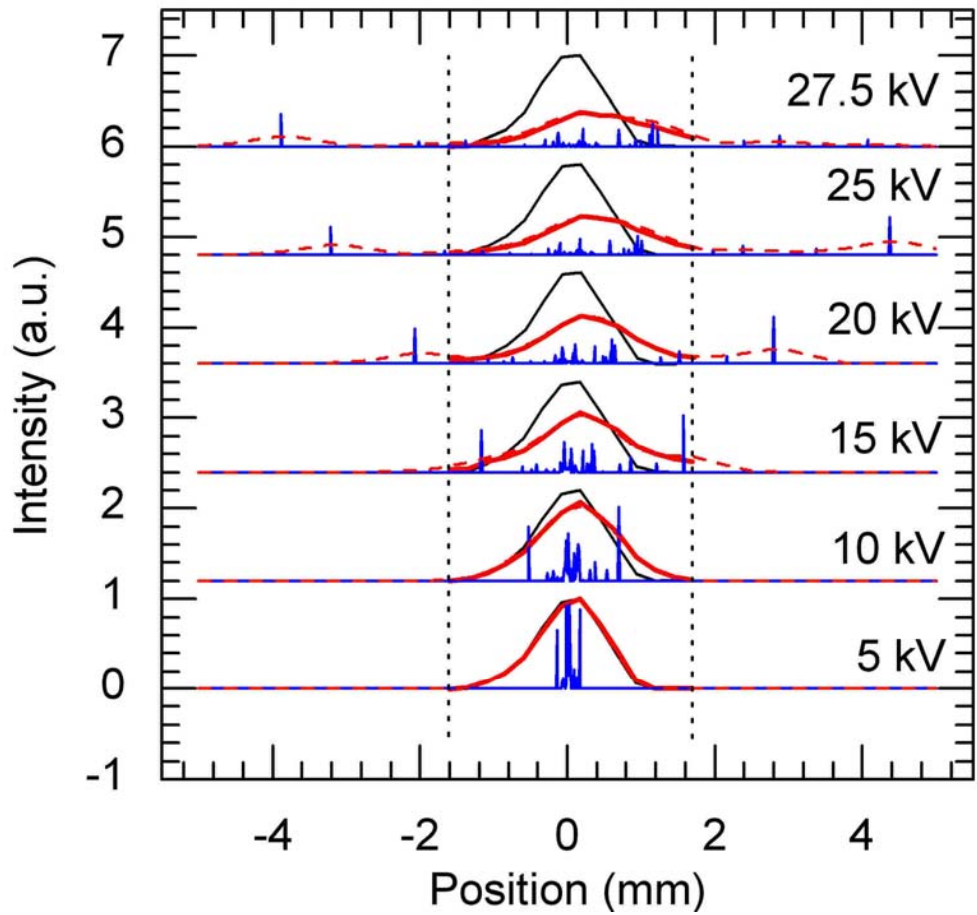

**Figure 4**

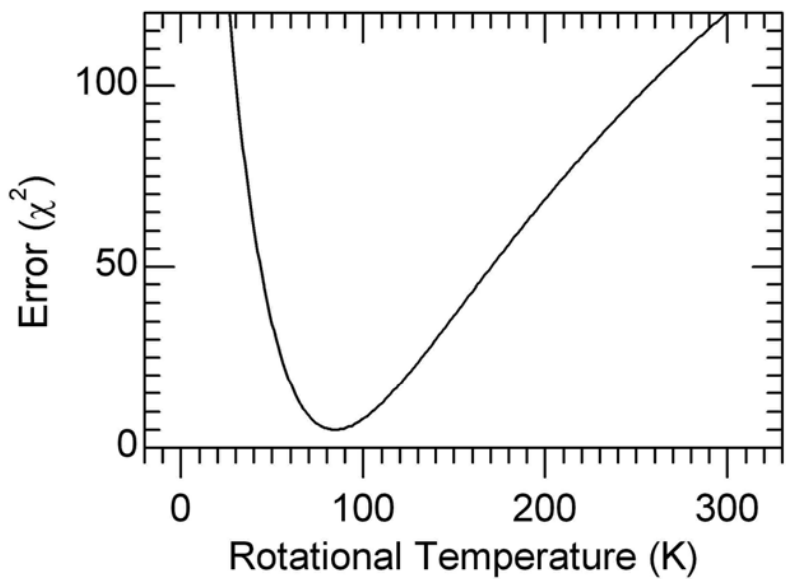

**Figure 5**

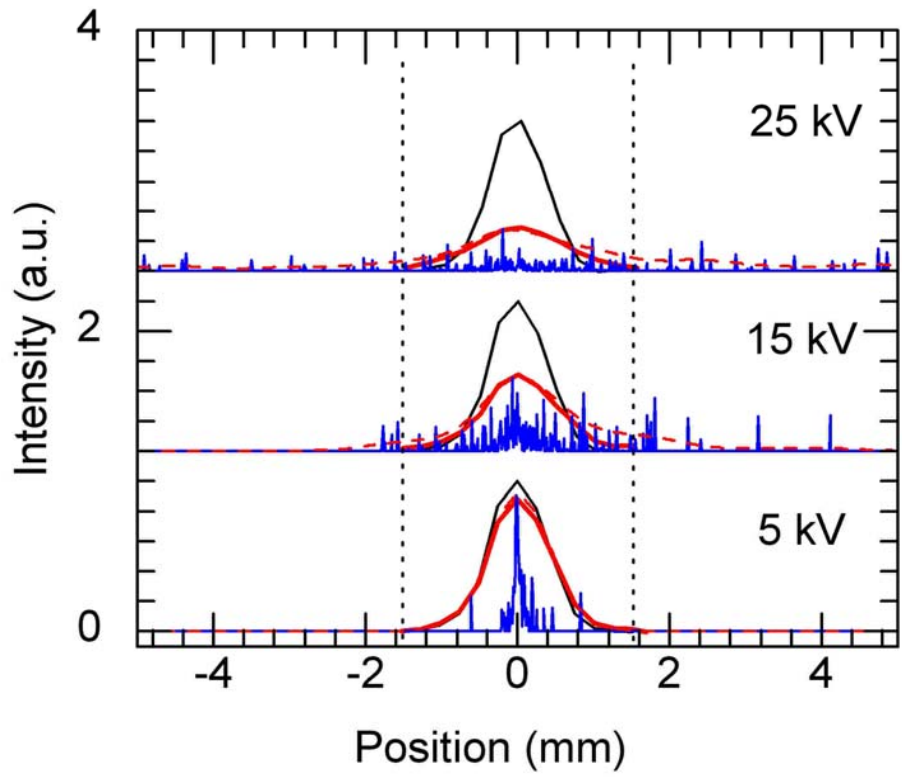

**Figure 6**

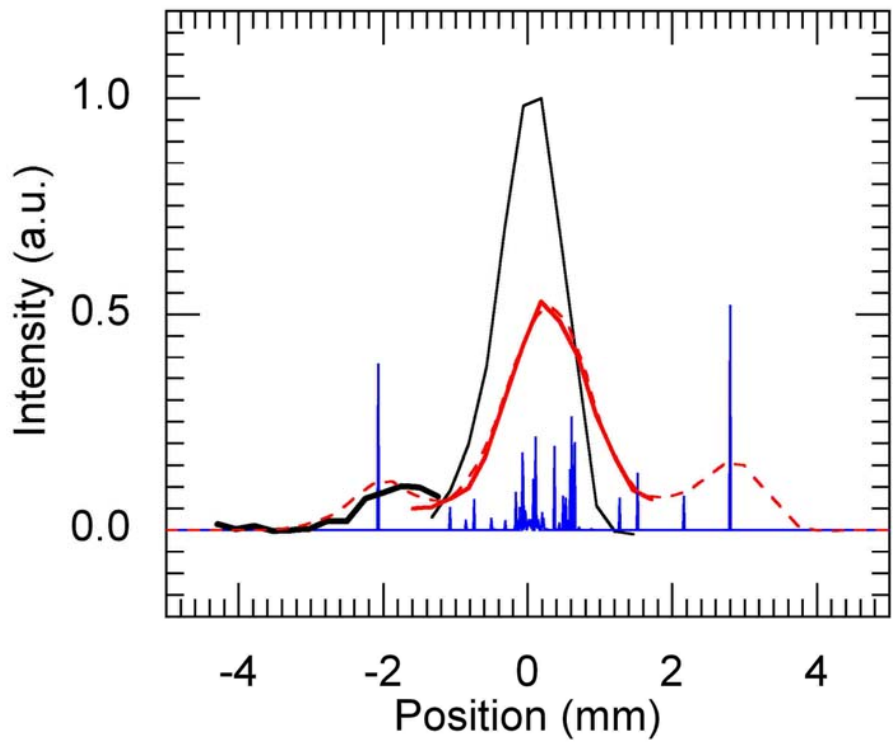

**Figure 7**

# REFERENCES

[1] S. A. Clough, Y. Beers, G. P. Klein, and L. S. Rothman, J. Chem. Phys. **59**, 2254 (1973).

[2] G. Herzberg, *Molecular Spectra and Molecular Structure*, Part II (Van Nostrand Reinhold, New York, 1945), p. 488.

[3] H. Scheffers, Phys. Z. **40**, 1 (1939).

[4] R. G. J. Fraser, *Molecular Beams* (Chemical Publishing Co., New York, 1938).

[5] R. Moro, R. Rabinovitch, C. Xia, and V. V. Kresin, Phys. Rev. Lett. **97**, 123401 (2006).

[6] N. F. Ramsey, *Molecular Beams* (Oxford University Press, Oxford, 1956).

[7] G. Tikhonov, K. Wong, V. Kasperovich, and V. V. Kresin, Rev. Sci. Instrum. **73**, 1204 (2002).

[8] NIST Mass Spec Data Center, "Mass Spectra," in NIST Chemistry WebBook, NIST Standard Reference Database No. 69, ed. by P. Linstrom and W. G. Mallard (National Institute of Standards and Technology, Gaithersburg, 2003), http://webbook.nist.gov/chemistry.

[9] W. S. Benedict, N. Gailar, and E. K. Pleyler, J. Chem. Phys. **21**, 1301 (1953).

[10] A. H. Brittain, A. P. Cox, G. Duxbury, T. G. Hersey, and R. G. Jones, Mol. Phys. **24**, 843 (1972).

[11] J. Bulthuis, J. Möller, and H. J. Loesch, J. Phys. Chem. A **101**, 7684 (1997).

[12] C. H. Townes and A. L. Schawlow, *Microwave Spectroscopy* (Dover, New York, 1975).

[13] J. M. Flaud, C. Camy-Peyret, and J. P. Maillard, Mol. Phys. **32**, 499 (1976).

[14] G. Avila, J. Chem. Phys. **122**, 144310 (2005).

[15] M. A. Omar, *Elementary Solid State Physics* (Addison-Wesley, Reading, 1993).

[16] M. Schnell, C. Herwig, and J. A. Becker, Z. Phys. Chem. **217**, 1003 (2003).

[17] V. Visuthikraisee and G. F. Bertsch, Phys. Rev. A **54**, 5104 (1996).

[18] M. Abd El Rahim, R. Antoine, M. Broyer, D. Rayane, and Ph. Dugourd, J. Phys. Chem. A **109**, 8507 (2005).

[19] R. Antoine, M. Abd El Rahim, M. Broyer, D. Rayane, and Ph. Dugourd, J. Phys. Chem. A **110**, 10006 (2006).